\documentclass[aps,prl,twocolumn,superscriptaddress, longbibliography]{revtex4-2}
\usepackage{gensymb}
\usepackage{color}
\usepackage[normalem]{ulem}
\usepackage{graphicx}% Include figure files
\usepackage{dcolumn}% Align table columns on decimal point
\usepackage{float}
\usepackage{bm}% bold math % physical review styleguide says: bold three-vector notation (e.g., k) not italic
\usepackage{amsmath}
\usepackage{amsfonts}
\usepackage{comment}

\usepackage{hyperref}

\newcommand{\vb}{\mathbf}

\begin{document}
\title{
Symmetry-Enforced Nodal \textit{f}-Wave Magnets
}

\author{Moritz M. Hirschmann}
\thanks{moritz.hirschmann@riken.jp}
\affiliation{RIKEN Center for Emergent Matter Science (CEMS), Wako, Saitama 351-0198, Japan}

\author{Akira Furusaki}
\affiliation{RIKEN Center for Emergent Matter Science (CEMS), Wako, Saitama 351-0198, Japan}

\author{Max Hirschberger}
\thanks{hirschberger@ap.t.u-tokyo.ac.jp}
\affiliation{RIKEN Center for Emergent Matter Science (CEMS), Wako, Saitama 351-0198, Japan}
\affiliation{Department of Applied Physics, The University of Tokyo, Bunkyo, Tokyo 113-8656, Japan}

\date{\today}

%% prl: "abstract not more than 600 characters including spaces", it seems no recently published prl is even remotely close to follow this rule, typically 1000+ characters even 1500 is possible, so probably it does not matter much as long is it is concise. 
\begin{abstract}
Owing to their relevance for spintronics, electronic band splitting and spin-polarization textures in magnets are active areas of research.
In non-collinear magnets, alternating spin textures can arise both for isolated bands and for intersecting band pairs with nodal splitting.
This raises the question of whether $p,f,\ldots$-wave magnets should be defined by their spin polarization or their band splitting. 
To resolve this ambiguity, we introduce spin-space symmetries that couple the spin polarization and splitting textures for all bands.
Focusing on the nodal $f$-wave magnet, we construct a tight-binding model of itinerant electrons on a honeycomb bilayer coupled to a non-collinear magnetic texture.
Analytic expressions for spin polarization and splitting reveal the dependence on hopping and exchange coupling.
We predict a canting-induced spin conductivity arising from the nodal structure of the splitting.
Furthermore, the $f$-wave magnet in the bulk can induce $p$-wave magnetism on the surface.
This surface $p$-wave character leads to a bulk-forbidden Edelstein effect with $f$-wave anisotropy.
\end{abstract}

\maketitle

%%%%%%%%%%%%%%%%%%%%%%%%%%%%%%%%%%%%%%%%%%%%%%%%%%%%%%%%%%%%%%%%%%%%%%%%%%%%%%%%
%limitation: 3,750 words, from the abstract to the acknowledgement, including figure caption
%Figure: up to 4

\textit{Introduction} -- Magnetic textures can induce an energy splitting of itinerant electrons.
Even in ferromagnets, this splitting is inhomogeneous \cite{ Hirschmann2022, lou2024orbital} because spin-orbit coupling (SOC) and orbital polarization vary across the Brillouin zone. 
Likewise, a spatially inhomogeneous magnetic texture may produce an anisotropic splitting and polarization textures in momentum space \cite{BohmJung2024, gurung2024nearly, liu2024unconventional}. 

\begin{figure}[htb]
  \begin{center}
		\includegraphics[width=0.99\linewidth]{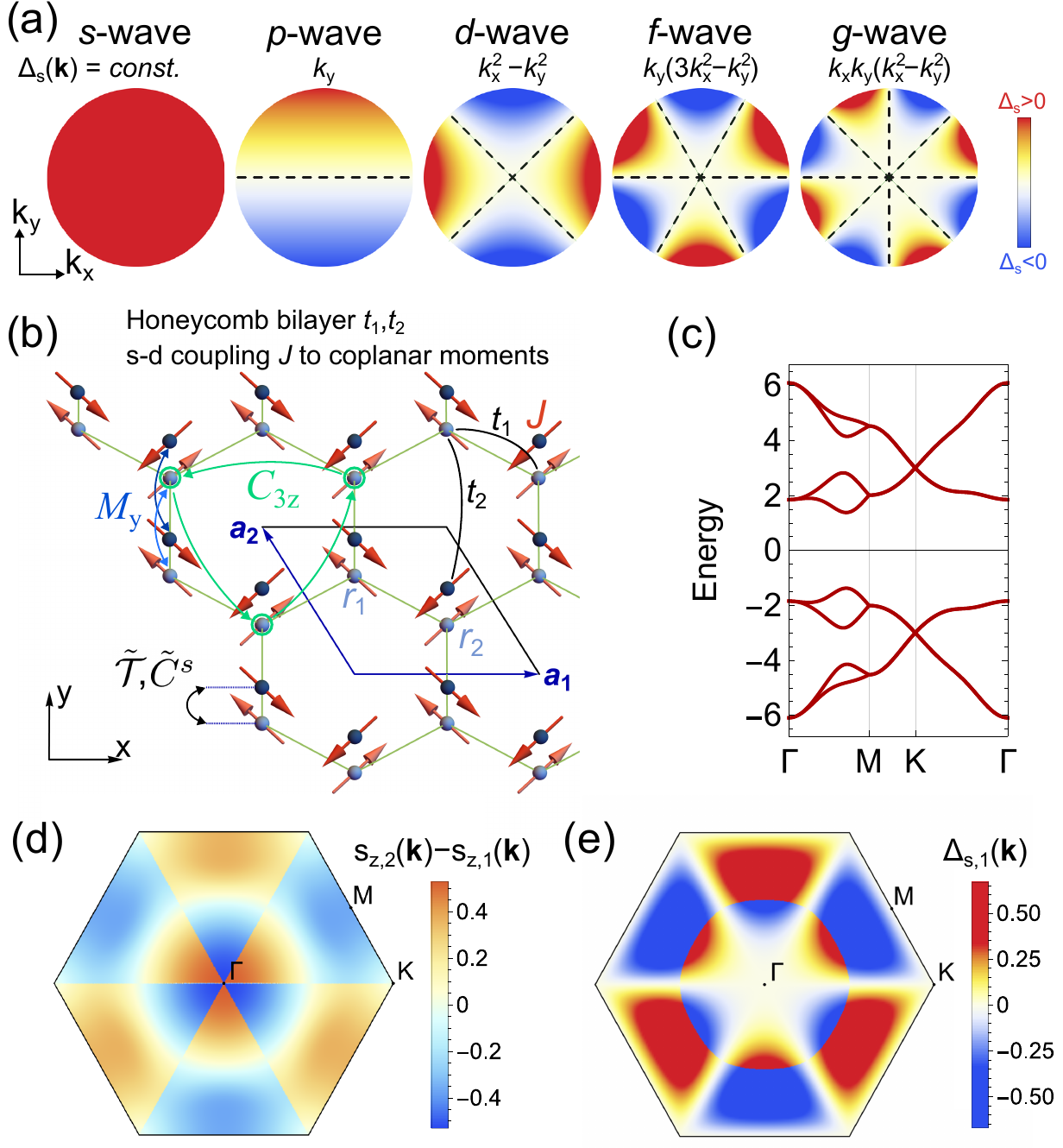}
    \caption[Spin splitting, spin polarization for an f-wave magnet]{
    Spin splitting, spin polarization for an $f$-wave magnet.
    (a) Comparison $s,p,d,f,g$-wave spin splitting.
    (b) Minimal $f$-wave magnetic texture with symmetries $\widetilde{C}^s, \widetilde{\mathcal{T}}, $ and $C_{3z}$. Top/bottom layer with dark-/light-colored spheres. 
    (c) Electronic band structure exhibiting spin-degenerate lines on M-K-$\Gamma$.
    (d) Difference of band-resolved spin $s_z$ polarization $s_{z,n}(\vb{k})$ between the first and second bands. 
    (e) Spin splitting $\Delta_{s,1}(\vb{k})$ between the first and second bands. $t_1 = 1$, $t_2 = 1/2$, $J = 3$.
    }
    \label{Fig1}
  \end{center}
\end{figure}

Momentum-dependent spin splitting can be classified by the presence of gapless points, lines, or planes.
%, which result from magnetic and spin-space groups determined by the magnetic texture and crystal structure.
For example, collinear altermagnets at negligible SOC exhibit an anisotropic spin splitting that vanishes in nodal lines in two dimensions (2D) or planes in three dimensions (3D) \cite{naka2019spin, Smejkal2022PRXBeyondConv, Smejkal2022PRXEmergingResearch}.
A collinear magnetic texture implies that the spin $s_z$ along the collinear order is a good quantum number at every momentum and determines the splitting structure.
For example, in a $d$-wave altermagnet, a spin-up band (red) can be mapped to a spin-down band (blue) by a fourfold rotation, leading to lines of vanishing splitting, $\Delta_s(\vb{k}) =0$, see Fig.~\ref{Fig1}(a).

To describe the relation between spin polarization and band splitting, one may define a splitting $\vert \Delta_{s,n}(\vb{k}) \vert \equiv  E_{n+1}(\vb{k}) - E_{n}(\vb{k})$ between energies of bands $n+1$ and $n$ that would be degenerate without magnetic exchange.
The spin polarization defines the sign of $\Delta_{s,n}(\vb{k})$.
For altermagnets, we define $\Delta_{s,n}(\vb{k}) > 0$ if the spin $s_z$ expectation value of band $n$ is negative, 
$s_{z,n}(\vb{k}) \equiv  \langle \psi_n(\vb{k})|s_z|\psi_n(\vb{k}) \rangle < 0$.
The spin splitting $\Delta_{s,n}(\vb{k})$ can be expanded in real spherical harmonics labeled by $l \in \mathbb{N}_0$.
Ferromagnets correspond to $s$-wave ($l=0$) and collinear altermagnets exhibit $d,g,i,\ldots$-wave ($l$ even) \cite{Smejkal2022PRXEmergingResearch}.
The parameter $l$ corresponds to the number of nodal lines in 2D (dashed lines in Fig.~\ref{Fig1}(a)) or of nodal planes in 3D.

To realize odd-parity magnets ($p,f,\ldots$), two types of systems have been considered: 
(I) Collinear magnets with broken spinless time-reversal symmetry by Floquet driving \cite{huang2025light, zhu2025floquet, Liu2026LightInducedOddParity} or orbital magnetism \cite{gonzalez2025model, zhuang2025oddparityOrbitalOrder, leeb2026collinear};
and (II) non-collinear magnets \cite{hellenes2023p}.
For magnets (II), unlike altermagnets, spin is not a good quantum number.
Instead, their spin polarization varies with momentum even at non-zero splitting $\Delta_{s,n}(\vb{k})$. 
For example, the spin polarization texture of a non-collinear magnet, Fig.~\ref{Fig1}(d), shows a continuous change from $\Gamma$ to M points. 
Nevertheless, non-collinear systems are used in experimental realizations of $p$-wave magnets \cite{yamada2025metallic, song2025electrical} and theoretical studies of odd-parity magnets \cite{hellenes2023p, chakraborty2025highly, uchino2025analysis, kudasov2025topological, okumura2025spiral, luo2025spin, Lee2026IncommensurationOddParity, mitscherling2026microscopicpWave, Yu2025prl, neumann2026antialtermagneticmagnons}.
Specifically, the splitting in $p$-wave magnets leads to anisotropic transport \cite{hellenes2023p, yamada2025metallic}, and if time-reversal symmetry is broken there is an anomalous Hall effect arising from the gapped nodes \cite{yamada2025metallic, okumura2025spiral}. 
Likewise, the $p$-wave Edelstein effect requires both spin polarization and band splitting \cite{chakraborty2025highly, ezawa2025out}.

How can we integrate non-collinear magnets into the $s,p,d,\ldots$-wave classification scheme shown in Fig.~\ref{Fig1}(a), where $\Delta_{s}(\vb{k})$ describes spin polarization and splitting? 
To answer this question, we focus on $f$-wave magnets, as they have so far mostly been studied in collinear magnets with fully polarized bands \cite{Ezawa2026quantumGeometryReview, Ezawa2025quanthall, Ezawa2025higherOrderResponses, zeng2025odd, huang2025light, zhu2025floquet, gonzalez2025model}.
Since the previously discussed non-collinear models \cite{uchino2025analysis, luo2025spin, kudasov2025topological, neumann2026antialtermagneticmagnons, Yu2025prl} do not exhibit an $f$-wave splitting for all bands, we present symmetry arguments that enforce $f$-wave splitting and polarization.
This symmetry-enforced low-energy description implies a characteristic bulk spin conductivity, if the order is canted, and an anisotropic surface Edelstein effect.

%%%%%%%%%%%%%%%%%%%%%%%%%%%%%%%%%%%%%%%%%%%%%%%%%%%%
%%%%%%%%%%%%%%%%%%%%%%%%%%%%%%%%%%%%%%%%%%%%%%%%%%%%

\textit{Symmetries of f-wave magnets} -- With a spin rotation $C^s_{z,n}$ of order $n \in \mathbb{N}$ around the $z$ axis, the $s_x$ and $s_y$ components of  spin polarization vanish (Eq.~\eqref{Eq_VanishingSpinPol}) \cite{calvo1978quantum, hellenes2023p, brekkePRL2024, yamada2025metallic}. 
A non-collinear texture may only be invariant under the combination of $C^s_{z,n}$ and a spatial operation, e.g., a translation in a magnetic spiral \cite{sandratskii1991symmetry}.
Here, we consider a bilayer (Fig.~\ref{Fig1}(b)) where the top and bottom layers are related by a mirror symmetry $M_z$, which only acts on space, not on spin.
We choose a magnetic texture that respects the composite symmetry $\widetilde{C}^s \equiv C^s_{z,2} M_z$.
The symmetry $\widetilde{C}^s$ implies that the spin polarization $s_{x,n}(\vb{k})$ of any band $n$ and momenta $\vb{k}$ vanishes,
\begin{align}
    s_{x,n}(\vb{k})
    &=  \langle \psi_n(\vb{k}) | \widetilde{C}^{s\dagger} \widetilde{C}^s s_x | \psi_n(\vb{k}) \rangle
    \nonumber
    \\
    &\!\!\!\!\!= - \langle \psi_n(\vb{k})| \widetilde{C}^{s\dagger} s_x \widetilde{C}^s | \psi_n(\vb{k}) \rangle
    = -s_{x,n}(\vb{k}) = 0,
    \label{Eq_VanishingSpinPol}
\end{align}
where we used the anticommutation $\{\widetilde{C}^s, s_{x} \} = 0$ and that all $\psi_n(\vb{k})$ are also eigenstates of $\widetilde{C}^s$.
While Eq.~\eqref{Eq_VanishingSpinPol} follows from the spin actions, $\widetilde{C}^s = i \sigma_z$ and $s_i = \tfrac{1}{2} \sigma_i$ with Pauli matrices $\sigma_i$ and $\hbar = 1$, it holds for any $n$ in $C^s_{z,n}$. 
Thus, $\widetilde{C}^s$-symmetry implies $s_{x,n}(\vb{k}) = s_{y,n}(\vb{k}) = 0$ and only $s_{z,n}(\vb{k}) \neq 0$ for all $\vb{k}$, which enables us to define the sign of spin splitting $\Delta_{s,n}(\vb{k})$ between bands $n$ and $n+1$ as \mbox{$\text{sgn}(\Delta_{s,n}(\vb{k})) \equiv \text{sgn}(s_{z,n+1}(\vb{k}) - s_{z,n}(\vb{k}))$} in analogy to altermagnets.

For the odd-parity spin splitting, $\Delta_s(\vb{k}) = -\Delta_s(-\vb{k})$, we consider symmetries including time reversal $\mathcal{T} = i \sigma_y K$ with complex conjugation $K$. 
Coplanar textures normal to the $z$ axis always exhibit $C^s_{z,2} \mathcal{T}$ \cite{Smejkal2022PRXEmergingResearch}. 
While this spin rotation with time reversal $C^s_{z,2} \mathcal{T}$ implies odd-parity spin polarization $s_{z,n}(\vb{k}) = - s_{z,n}(-\vb{k})$, it does not enforce a band degeneracy at $\vb{k} = \vb{0}$ \cite{uchino2025analysis} because $(C^s_{z,2} \mathcal{T})^2 = 1$. 
With a combined symmetry $\widetilde{\mathcal{T}} \equiv \mathcal{T} M_z$ comprising time reversal and the mirror, level crossings occur at invariant momenta for every band due to Kramers theorem because $(\mathcal{T} M_z)^2 =-1$.
This ensures an odd-parity spin splitting $\Delta_{s,n}(\vb{k}) = -\Delta_{s,n}(-\vb{k})$ (Fig.~\ref{Fig1}(e)).
If the system is 3D, the mirror $M_z$ in $\widetilde{\mathcal{T}}$ can be replaced by a translation in the $z$ direction.

In non-collinear magnets, $s_{z}$ is not a symmetry operator, and thus a spin polarization texture $s_{z,n}(\vb{k})$ alone does not protect band crossings.
%In addition to the odd-parity splitting, 
Nevertheless, the symmetries $\widetilde{C}^s$ and $\widetilde{\mathcal{T}}$ lead to lines (or surfaces in 3D) along which the spin splitting vanishes. 
These follow from the symmetry eigenvalues $\lambda^s_n(\vb{k}) = \langle \psi_n(\vb{k}) | \widetilde{C}^s | \psi_n(\vb{k}) \rangle$ of band~$n$. 
As $\widetilde{C}^s$ is momentum-independent, its eigenvalues $\lambda^s_n(\vb{k})$ are piece-wise constant in $\vb{k}$ and generally complex-valued $\lambda^s_n(\vb{k}) \notin \mathbb{R}$.
Time reversal $\widetilde{\mathcal{T}}$ (or $C^s_{\perp,2} \mathcal{T}$) implies $\lambda^s_n(\vb{k}) = \lambda^s_n(-\vb{k})^*$.
By an argument analogous to how time reversal-related mirror eigenvalues enforce almost movable nodal lines, any path connecting $-\vb{k}$ to $\vb{k}$ must cross at least one band degeneracy \cite{hirschmann2021symmetry, YangBJ2017, figgemeier2025imaging, domaine2025tunable}. 
Thus, each Kramers degeneracy is part of a nodal line comprising the eigenvalues $\lambda^s_n(\vb{k})$ and $\lambda^s_n(\vb{k})^*$.

The simplest spin-splitting in 2D that is consistent with $\widetilde{C}^s$ and $\widetilde{\mathcal{T}}$ exhibits a single nodal line ($p$-wave).
To enforce $f$-wave splitting with three nodal lines, Fig.~\ref{Fig1}(a), we introduce a threefold rotation symmetry $C_{3z}$ that commutes with $s_z$ and $\widetilde{C}^s$.
Mirror symmetries are optional, but can enforce straight nodal lines.
For example, the spinful mirror $M_y$ in Fig.~\ref{Fig1}(b) pins a nodal plane to $k_y = 0$ by enforcing $\Delta_s(k_y) = -\Delta_s(-k_y)$ and thus $\Delta_{s}(\vb{k}) \propto k_y(3k_x^2 - k_y^2)$ around $\vb{k} =0$.
Unlike the 2D case, in 3D it is possible that a threefold rotation maps a single nodal surface to itself, i.e., $C_{3z}$ is not sufficient to enforce $f$-wave splitting. 
Instead, three mirror operations that fix nodal surfaces are required to enforce $f$-wave in 3D.
Thus, four out of seven real $f$-wave spherical harmonics can be enforced, namely, $k_y(3k_x^2 - k_y^2), k_x(3k_y^2 - k_x^2), k_x k_y k_z,$ and $k_z(k_x^2-k_y^2)$.

In $f$-wave magnets, symmetries that enforce spin degeneracy at all $\vb{k}$ and even-parity spin polarization must be absent.

%%%%%%%%%%%%%%%%%%%%%%%%%%%%%%%%%%%%%%%%%%%%%%%%%%%%
%%%%%%%%%%%%%%%%%%%%%%%%%%%%%%%%%%%%%%%%%%%%%%%%%%%%

\textit{Tight-binding model} -- We construct a minimal 2D Hamiltonian $H(\vb{k})$ for electrons in an $f$-wave magnet that is symmetric under $\widetilde{C}^s$, $\widetilde{\mathcal{T}}$,  $C_{3z}$, and $M_y$, see Fig.~\ref{Fig1}(b). 
The nearest-neighbor hopping $t_1$ stays within each honeycomb layer and a long-range hopping $t_2$ joins adjacent layers.
If $t_1 = 0$ or $t_2=0$, the $f$-wave splitting vanishes in our model.
The magnetic unit cell comprises 4 sites, 
each with an s-d coupling $J$ between the electron and a local magnetic moment. 
Once one of the coplanar moments is chosen, symmetries $\widetilde{C}^s$ and $M_y$ fix the remaining moments. 
To avoid a collinear magnetic texture, the chosen moment may not be along the $x$ or $y$ axis; we choose the moment at $\vb{r}_1$ along $(-1,1,0)$.
This model can be extended to 3D, see \hyperref[EndMatter_TBmodell]{End Matter}.

The band structure, Fig.~\ref{Fig1}(c), exhibits the expected splitting with nodal lines along $\Gamma$-K and K-M, as well as Dirac points at $K$ inherited from the honeycomb lattice \cite{castro2009electronic}. 
In the radial direction of Fig.~\ref{Fig1}(d), a sign change in the spin polarization appears at the avoided crossing at energy $E = \pm 3.5$.
Sign changes of polarization without a band degeneracy are common in odd-parity magnets \cite{hellenes2023p, brekkePRL2024, uchino2025analysis}.
While the splitting $\Delta_{s}(\vb{k})$ has $f$-wave character for small $\vb{k}$ in Fig.~\ref{Fig1}(e), the sign change in spin polarization leads to a sign change of $\Delta_{s}(\vb{k})$ in the radial direction.

%%%%%%%%%%%%%%%%%%%%%%%%%%%%%%%%%%%%%%%%%%%%%%%%%%%%
%%%%%%%%%%%%%%%%%%%%%%%%%%%%%%%%%%%%%%%%%%%%%%%%%%%%

\textit{Quantifying f-wave splitting} -- In the eigenbasis of $\widetilde{C}^s$ the two-band continuum expansion $H_{\Gamma}(\vb{k})$ around the $\Gamma$ point of the $f$-wave model $H(\vb{k})$ is diagonal. 
The symmetries $\widetilde{\mathcal{T}}$ and $C_{3z}$ restrict the lowest order terms to
\begin{align}
    H_{\Gamma}(\vb{k})
    = 
    E_0^{(1)} \,  (k_x^2 + k_y^2) \tilde{\sigma}_0 
    + \Delta_s^{(1)} \,a k_y (3 k_x^2 - k_y^2) \tilde{\sigma}_z,
    \label{Eq_lowEHam}
\end{align}
where $E_0^{(1)}, \Delta_s^{(1)} \in \mathbb{R}$ are the lowest-order expansion coefficients corresponding to the non-magnetic band dispersion and the size of the $f$-wave splitting, respectively. 
In the following, we set the lattice constant $a =1$. 
The matrices $\tilde{\sigma}_0$ and $ \tilde{\sigma}_i$ are the identity and Pauli matrices in the eigenbasis at $\Gamma$.
In this basis, the spin operator $s_z$ is $\vert s_{z,n}(\vb{k} \rightarrow \vb{0}) \vert \tilde{\sigma}_z$. 

\begin{figure}[t]
  \begin{center}
		\includegraphics[width=0.99\linewidth]{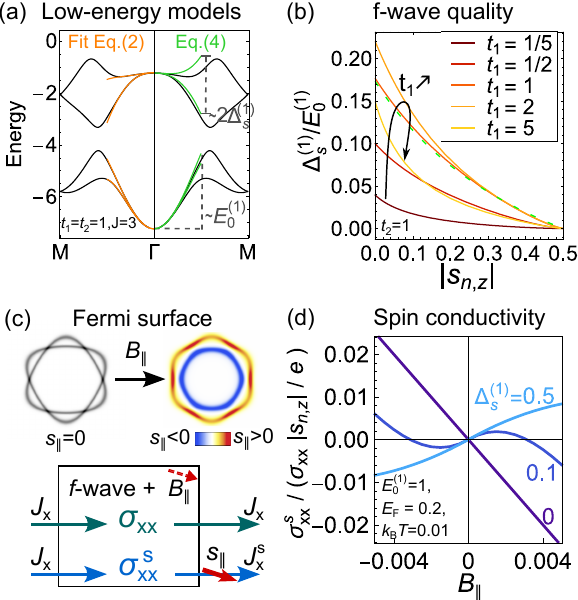}
    \caption[Origin of f-wave splitting]{
    Origin of $f$-wave splitting.
    (a) Comparison of numerical fit (orange) up to $k^5$ and the analytical low-energy expansion (green).
    (b) $f$-wave splitting divided by size of parabolic part as function of spin polarization for the lower band pair. 
    The black arrow highlights the effect of increasing hopping $t_1$. 
    The green dashed line is the analytical result for $t_1 = t_2$.
    (c) An in-plane field $B_\parallel$ gaps the nodal points on the Fermi surfaces and creates a polarization $s_\parallel$.
    (d) The spin conductivity $\sigma^s_{xx}$ divided by electric conductivity $\sigma_{xx}$ and spin polarization $\vert s_{n,k} \vert$ depends qualitatively on the $f$-wave splitting $\Delta_s^{(1)}$.
    }
    \label{Fig2}
  \end{center}
\end{figure}

Such an expansion matches the tight-binding model, orange line in Fig.~\ref{Fig2}(a). 
The $f$-wave nodal line structure fixes the second term in Eq.~\eqref{Eq_lowEHam} to $\propto k_y (3 k_x^2 - k_y^2)$. 
Thus, the symmetries are sufficient to enforce $f$-wave splitting by suppressing any $p$-wave terms.

To understand the origin of $f$-wave splitting in terms of hopping and exchange coupling, for $t_1 = t_2 = t > 0$ we expand the Hamiltonian $H(\vb{k})$ around $\vb{k} = 0$, see  \hyperref[EndMatter_LowEnergyModel]{End Matter}. 
The spin polarization $s_z$ for the basis states of Eq.~\eqref{Eq_lowEHam} at $\Gamma$ is independent of band index $n$ 
\begin{align}
    \vert s_{n,z} \vert = \frac{3 t}{2 \sqrt{J^2 + 9 t^2}}.
    \label{Eq_SpinPol}
\end{align}
The spin polarization increases with hopping $t$, and decreases with the exchange coupling $J$.
This $J$ dependence follows from the orthogonality of electron polarization $s_z$ and the coplanar moments of the magnetic texture.
The analytic expressions for the parabolic $E_0^{(1)}$ and $f$-wave splitting $\Delta_s^{(1)}$ are 
\begin{align}
    E_0^{(1)}
    \!&=\!
    \tfrac{5}{8} t \left( 1 + 2 \vert s_{n,z} \vert \right) , 
    \quad
     \Delta_s^{(1)}
    \!=\! \tfrac{\sqrt{3}}{8} t \left( \tfrac{1}{2} - \vert s_{n,z} \vert \right). 
    \label{Eq_AnalyticCoefficients}
\end{align}
The parabolic term with $E_0^{(1)}$ is nearly linear in $t$ and weakly affected by $J$.
Interestingly, the band splitting coefficient $\Delta_s^{(1)}$ vanishes if either $t = 0$ or $J=0$.

As we will show, for the Edelstein effect, $\Delta_s^{(1)}/E_0^{(1)} \gg 1$ and large $s_{n,z} \approx 1/2$ are favorable.
Here, these quantities inversely correlate and $\Delta_s^{(1)}/E_0^{(1)} < 1$ for all $s_{n,z}$, Fig.~\ref{Fig2}(b).
Conversely, for the upper band pair in Fig.~\ref{Fig2}(a) the hybridization between adjacent bands leads to \mbox{$E_0^{(1)} \approx 0$} implying $\Delta_s^{(1)}/E_0^{(1)} \gg 1$, cf.\  Fig.~\ref{FigSuppl_S1}.

\textit{Spin conductivity} -- $f$-wave magnets exhibit no spin current response to linear order in the electric field $E_j$ \cite{Ezawa2025higherOrderResponses, ezawa2026nonlinear, uchino2025analysis}.
To study spin currents in the context of a canted $f$-wave texture, we add a small in-plane field $H_B = - B_\parallel \tilde{\sigma}_x$ to the low-energy model Eq.~\eqref{Eq_lowEHam} (Fig.~\ref{Fig2}(c)).
The result is a \emph{linear} spin conductivity $\sigma^s_{xx}$ \cite{Zelse2017}, defined for an in-plane-polarized spin current $j^s_i = \sigma^s_{ij} E_j$.
When  $\Delta_s^{(1)}$ is large, the $B_\parallel$-induced polarization $s_\parallel$ is concentrated at the nodes where $\Delta_s(\vb{k}) = 0$.
This reduces $\vert\sigma^s_{xx}\vert$ as expected.
Furthermore, the splitting-induced asymmetry of Fermi surfaces enables a sign change relative to the response of a parabolic band (Fig.~\ref{Fig2}(d)), cf.\ $p, d, f, g$-wave in Fig.~\ref{FigSuppl_S2}.

%%%%%%%%%%%%%%%%%%%%%%%%%%%%%%%%%%%%%%%%%%%%%%%%%%%%
%%%%%%%%%%%%%%%%%%%%%%%%%%%%%%%%%%%%%%%%%%%%%%%%%%%%

\textit{$f$-wave symmetry constrained Edelstein effect} -- 
The Edelstein effect refers to a non-zero net spin $(\delta s)_i$ caused by an applied electric field $E_j$.  
Symmetry constraints on the Edelstein tensor $\chi_{ij}$, defined as $(\delta s)_i = \chi_{ij} E_j$, follow from Neumann's principle: Symmetry transformations leave the tensor invariant \cite{Watanabe2018symmetries, cracknell1969group}. 
For $p$ and $f$-wave magnets, the spin rotation $\widetilde{C}^s$ removes the rows of two spin components, $\chi_{xj} = \chi_{yj} = 0$ for any $j \in \{x,y,z\}$. 
Non-vanishing $\chi_{zj}$ components were calculated for $p$-wave magnets  \cite{chakraborty2025highly, ezawa2025out}.
For $f$-wave magnets, the three-fold rotation $C_{3z}$ conserves the remaining spin component but rotates the spatial components, setting $\chi_{zj} = 0$ for $j=x,y$, leaving only $\chi_{zz}$.
However, if additionally a mirror symmetry sends $s_z \rightarrow - s_z$ and $z \rightarrow z$, as for $\Delta_s(\vb{k}) \propto k_x(3k_y^2 - k_x^2)$ and $\propto k_y(3k_x^2 - k_y^2)$, then $\chi_{zz} = 0$.
In conclusion, the Edelstein effect in our $f$-wave model is forbidden in the bulk.

%%%%%%%%%%%%%%%%%%%%%%%%%%%%%%%%%%%%%%%%%%%%%%%%%%%%
%%%%%%%%%%%%%%%%%%%%%%%%%%%%%%%%%%%%%%%%%%%%%%%%%%%%

\begin{figure}[htb]
  \begin{center}
		\includegraphics[width=0.99\linewidth]{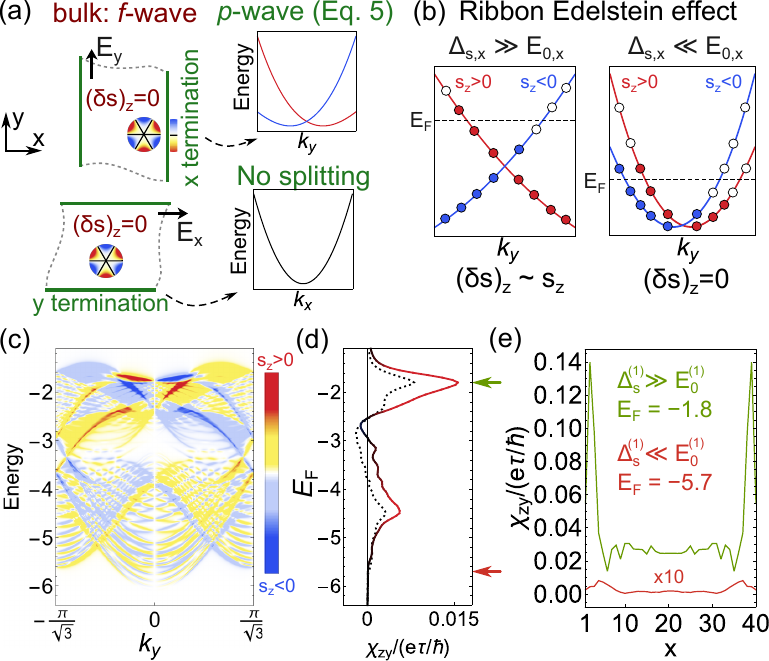}
    \caption[Anisotropic, edge Edelstein effect]{
    Anisotropic Edelstein effect on $f$-wave magnet ribbons. 
    (a) Ribbon geometry in $x$ termination leads to $p$-wave splitting, whereas $y$ termination does not.  
    (b) Schematic Edelstein effect for different splitting, where filled circles denote occupied states.
    (c) Band structure of an x-terminated ribbon, spin polarization shown in color.
    (d) Edelstein susceptibility $\chi_{zy}$ for the ribbon in (c); solid (dashed) lines denote the contribution from edge (bulk) sites at a filling $E_\text{F}$. 
    (e) Edelstein effect over position $x$ on the ribbon at Fermi energies marked in (d). $t_1 = 1,\, t_2 = 0.5, \, J = 3$. Ribbon width: 40 unit cells
}
    \label{Fig3}
  \end{center}
\end{figure}

\begin{figure}[t]
  \begin{center}
		\includegraphics[width=0.99\linewidth]{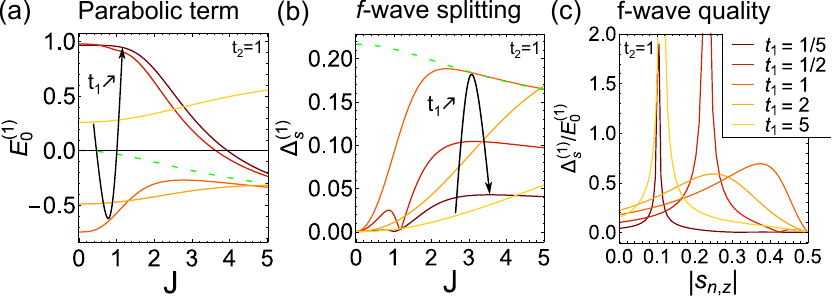}
    \caption[Origin of f-wave splitting]{
    Parameter dependence of $f$-wave splitting for the upper band pair in Fig.~\ref{Fig2}(a).
    (a) Parabolic dispersion and (b)  $f$-wave splitting as function of exchange coupling $J$. (c) Ratio of (a,b) over spin polarization. 
    Black arrows qualitatively highlight the effect of increasing hopping $t_1$. 
    The green dashed line is the analytical result for $t_1 = t_2$, which does not match the fitted parameters for small $J$ where the two-band approximation fails.
    Due to the band hybridization, the curves (a-c) are non-monotonous, unlike Fig.~\ref{Fig2}(a).
    Vanishing $E_0^{(1)}$ leads to divergences in (c).
    }
    \label{FigSuppl_S1}
  \end{center}
\end{figure}

\begin{figure}[t]
  \begin{center}
		\includegraphics[width=0.99\linewidth]{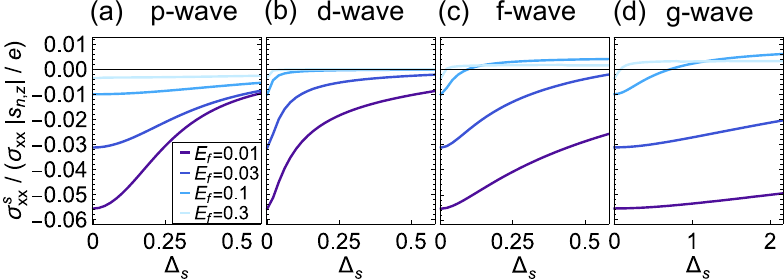}
    \caption[Linear spin current]{
    Normalized linear spin conductivity, Eqs.~\eqref{Eq_BoltzSigXX} and \eqref{Eq_BoltzSpinSigXX}, as a function of splitting $\Delta_s$ and Fermi energy $E_f$ for $B_\parallel = 0.001$ and $k_BT = 0.01$.
    The Hamiltonians are $k_x^2 + k_y^2 + \Delta_s(\vb{k}) \tilde{\sigma}_z$, where the splitting $\Delta_s(\vb{k})$ is defined in Fig.~\ref{Fig1}(a) with $\Delta_s$ as additional prefactor. 
    }
    \label{FigSuppl_S2}
  \end{center}
\end{figure}

\textit{$p$-wave from $f$-wave in finite systems} -- 
In a finite system, the crystal symmetries are reduced and two contributions to the Edelstein effect become allowed: 
(i) Finite-size effects prevent the cancellation due to $C_{3z}$ leading to bulk contributions, and (ii) surface effects that give a size-independent contribution. 
While any termination in the $x$-$y$-plane breaks three-fold rotation symmetry, $\widetilde{T}$ and $\widetilde{C}_s$ are preserved and lead to Kramers pairs at momentum $k = 0$ along the ribbon with a $p$-wave splitting for $k \neq 0$. 
For a ribbon with $x$-termination, the effective model for any Kramers pair at $\Gamma$ is given by
\begin{align}
    H_{x}(k_y) \equiv E_{0,x} k_y^2 + \Delta_{s,x} k_y \tilde{\sigma}_z,
    \label{Eq_pWaveEdge}
\end{align}
where $E_{0,x}, \Delta_{s,x} \in \mathbb{R}$, see Fig.~\ref{Fig3}(a).
Note, the transverse momentum $k_x$ in Eq.~\eqref{Eq_lowEHam} effectively takes only discrete values in a ribbon, bands overlap and hybridize.
Thus, Eq.~\eqref{Eq_pWaveEdge} describes only a small part of the ribbon bands.

A $y$-terminated ribbon (and symmetry-equivalent terminations) differs because $M_y$ sends $k_x \rightarrow k_x$ but $s_z \rightarrow -s_z$ implying $s_z(k_x) =0$.
Combined with $\widetilde{C}_s$ forcing $s_x(k_x) = s_y(k_x) = 0$, the bands have no spin texture and thus cannot exhibit $p$-wave splitting or an Edelstein effect (Fig.~\ref{Fig3}(a)).

%%%%%%%%%%%%%%%%%%%%%%%%%%%%%%%%%%%%%%%%%%%%%%%%%%%%
%%%%%%%%%%%%%%%%%%%%%%%%%%%%%%%%%%%%%%%%%%%%%%%%%%%%

\textit{Surface Edelstein effect} -- To study the Edelstein effect \cite{suzuki2023spin}, we use the Boltzmann equation in the relaxation time approximation \cite{johansson2024theory,  Ezawa2025higherOrderResponses, ezawa2025out}.
For a single band in a ribbon, Eq.~\eqref{Eq_pWaveEdge}, the non-zero component of the Edelstein effect, $(\delta s)_z = \chi_{zy} E_y$, can be analytically calculated at zero temperature.
The low-energy model assumes a constant spin polarization $\pm s_{z}$ for its basis states. 
As we have seen for the bulk in Fig.~\ref{Fig1}(d), the spin polarization changes with momentum, thus also the model Eq.~\eqref{Eq_pWaveEdge} only applies in a finite range of $k_y$.
This leads to two qualitatively different limits as illustrated in Fig.~\ref{Fig3}(b): (I) Dominant spin splitting $\Delta_{s,x} \gg E_{0,x}$ with two Fermi surfaces, yielding $\chi_{zy} = \tfrac{e \tau}{\pi \hbar} s_{z}$; (II) Dominant parabolic dispersion $\Delta_{s,x} \ll E_{0,x}$ with four Fermi surfaces and pairwise cancellation, leading to $\chi_{zy} = 0$.

%%%%%%%%%%%%%%%%%%%%%%%%%%%%%%%%%%%%%%%%%%%%%%%%%%%%
%%%%%%%%%%%%%%%%%%%%%%%%%%%%%%%%%%%%%%%%%%%%%%%%%%%%

\textit{Edelstein effect in the lattice model} -- We extend our discussion to full ribbon band structures using the tight-binding model $H(\vb{k})$.
As expected, for the $y$-terminated ribbon, the spin polarization $s_z(k_x)$ vanishes at all $k_x$.
For $x$ termination, there is an odd-parity spin polarization, Fig.~\ref{Fig3}(c).
Notably, the spin polarizations cancel where the bulk splitting is weak, at $E \approx -6$.

We calculated the Edelstein susceptibility $\chi_{zy}$ numerically as function of the filling $E_F$, Fig.~\ref{Fig3}(d).
As in our qualitative discussion, Fig.~\ref{Fig3}(b), the surface Edelstein effect is strongest where the bulk exhibits the largest $f$-wave splitting, i.e., at $E_F\approx-1.8$.
The spatially resolved contribution to the Edelstein effect (Fig.~\ref{Fig3}(e)) shows a finite-size effect in the bulk as well as a peaked surface contribution of comparable total magnitude (Fig.~\ref{Fig3}(d)).

%%%%%%%%%%%%%%%%%%%%%%%%%%%%%%%%%%%%%%%%%%%%%%%%%%%%
%%%%%%%%%%%%%%%%%%%%%%%%%%%%%%%%%%%%%%%%%%%%%%%%%%%%

%%% Discussion %%% 
\textit{Discussion} -- Since the surface Edelstein effect is a consequence of symmetry reduction, it can be generalized to $h$- and $j$-wave magnets. 
Magneto-optical spectroscopy \cite{Stamm2017MOKEspinhall, puebla2017direct}, or spin-sensitive transport measurements \cite{valenzuela2006direct, ghiasi2019charge, nomura2015temperature, Junyeon2017} may detect such a spin accumulation. 

In quasi-2D systems, an $f$-wave magnet can be realized even with SOC. 
A spinful mirror symmetry $M_z$ sets the polarizations $s_{x,n}(\vb{k}) = s_{y,n}(\vb{k}) = 0$ at all momenta. 
For example, the magnetic layer group $P\bar{6}m2$ (or $P_c\bar{6}m2$ with time reversal) leads to an $f$-wave magnet. 
While a non-collinear magnetic texture with this symmetry requires at least a sixfold Wyckoff position, with SOC, already Bernal bilayer graphene can exhibit $f$-wave spin polarization \cite{Koh2024}. 

We have shown that symmetry-enforced $p,f,\ldots$-wave splitting in non-collinear magnets matches the spherical harmonic classification of altermagnets \cite{Smejkal2022PRXBeyondConv, Ezawa2025higherOrderResponses} (Fig.~\ref{Fig1}(a)). 
With linear spin conductivity and $p$-wave splitting on the edges, we identified characteristic properties that arise from \emph{nodal} $f$-wave magnets.
The surface Edelstein effect, on the other hand, is also symmetry-allowed for $f$-wave polarization $s_{z,n}(\vb{k}) \propto k_y (3k_x^2 - k_y^2)$ without splitting.

%%%%%%%%%%%%%%%%%%%%%%%%%%%%%%%%%%%%%%%%%%%%%%%%%%%%
%%%%%%%%%%%%%%%%%%%%%%%%%%%%%%%%%%%%%%%%%%%%%%%%%%%%

\vskip\baselineskip
\textbf{Acknowledgements}\\
The authors are grateful to Ryota Nakano, Max T. Birch, Rinsuke Yamada, Alexander Mook and Hitoshi Seo for insightful discussions. 
M.~M.~H.~is funded by the RIKEN Special Postdoctoral Researcher Program.

\bibliography{OddParity_fwave_bibliography}

%%%%%%%%%%%%%%%%%%%%%%%%%%%%%%%%%%
%%%%%%%%%%%%%%%%%%%%%%%%%%%%%%%%%%
%% SUPPLEMENTARY INFORMATION
%%%%%%%%%%%%%%%%%%%%%%%%%%%%%%%%%%
%%%%%%%%%%%%%%%%%%%%%%%%%%%%%%%%%%
\clearpage

\renewcommand\theequation{S\arabic{equation}}
\setcounter{equation}{0} 

\renewcommand{\thetable}{S\arabic{table}}

\renewcommand\thesection{S \Roman{section}} 
\setcounter{section}{0}

\section*{End Matter}
% End Matter since 2024, example: https://journals.aps.org/prl/abstract/10.1103/PhysRevLett.134.157001 on the webpage it is called appendices, in the pdf it is called end matter and formated as the main text.

\textit{Tight-binding model} \label{EndMatter_TBmodell} -- The primitive basis vectors of the magnetic unit cell, see Fig.~\ref{Fig1}(b), are 
\begin{align}
    \vb{a}_1 = (1,0,0)^T, 
    \vb{a}_2 = \frac{1}{2}(- 1,\sqrt{3},0)^T, 
    \vb{a}_3 = (0,0,1)^T, 
\end{align}
where $\vb{a}_3$ becomes a translation vector in the 3D generalization of the discussed 2D model. 
Lattice sites of the paramagnetic Wyckoff position are given by
\begin{align}
    \vb{r}_1 = \frac{1}{3} \vb{a}_1 + \frac{2}{3} \vb{a}_2, \text{ and }
    \vb{r}_2 = \frac{2}{3} \vb{a}_1 + \frac{1}{3} \vb{a}_2.
\end{align}
The vector $\vb{t}_{1/2} = \vb{a}_3/2$ relates the sublattices: $\vb{r}'_1 = \vb{r}_1 + \vb{t}_{1/2}$ and $\vb{r}'_2 = \vb{r}_2 + \vb{t}_{1/2}$, which corresponds to AA stacked honeycomb layers. 
Note, to create an $f$-wave model, any other stacking that preserves the three symmetries $\widetilde{C}^s, \widetilde{\mathcal{T}}, $ and $C_{3z}$ is also possible. 

We first define the non-zero entries of the spinless $4 \times 4$ Hamiltonian $H_{0}(\vb{k})$. 
The basis is $(c_{1\uparrow}, c_{2\uparrow},c'_{1\uparrow},
c'_{2\uparrow})$ corresponding to the sites at $\vb{r}_1, \vb{r}_2, \vb{r}'_1, \vb{r}'_2$, respectively. 
We include the minimal number of terms to achieve $f$-wave splitting in 2D,
\begin{align}
    &(H_{0}(\vb{k}))_{12}
    = 
    (H_{0}(\vb{k}))_{34} 
    \nonumber\\
    &= 
    t_1 \left(
    \mathrm{e}^{i \Delta\vb{r} \cdot \vb{k}}
    + \mathrm{e}^{i (\Delta\vb{r} + \vb{a}_1 ) \cdot \vb{k}}
    + \mathrm{e}^{i (\Delta\vb{r} - \vb{a}_2) \cdot \vb{k}}
    \right)
    \\[.2cm]
    &(H_{0}(\vb{k}))_{14}
    = 
    (H_{0}(- \vb{k}))_{23}
    \nonumber\\
    &= 
    t_2 
    \left(
    \mathrm{e}^{i (\Delta\vb{r} + \vb{a}_1 - \vb{a}_2) \cdot \vb{k}} 
    + \mathrm{e}^{i (\Delta\vb{r} - \vb{a}_1 - \vb{a}_2) \cdot \vb{k}} 
    + \mathrm{e}^{i (\Delta\vb{r} + \vb{a}_1 + \vb{a}_2) \cdot \vb{k}} 
    \right)
\end{align}
where $\Delta\vb{r} = \vb{r}_1 - \vb{r}_2$, $t_1, t_2 \in \mathbb{R}$, and the remaining matrix form the hermiticity.
Note for the 2D model $\vb{t}_{1/2}$ and $\vb{a}_3$ are not translation vectors.

Next, we add the spin $\downarrow$  states to our basis, $(
c_{1\uparrow},c_{1\downarrow},
c_{2\uparrow},c_{2\downarrow},
c'_{1\uparrow},c'_{1\downarrow},
c'_{2\uparrow},c'_{2\downarrow})$. 
The full $f$-wave tight-binding model is the sum of the spinless Hamiltonian  and the s-d exchange Hamiltonian $H_{\text{spin}}$,
\begin{align}
    H(\vb{k}) = H_{0}(\vb{k}) \otimes \sigma_0 + H_{\text{spin}}, \label{Eq_fullTBham}
\end{align}
with the identity $\sigma_0$ in spin-space.
The spin texture is introduced as on-site spin splitting given by an $8 \times 8$ matrix,
\begin{align}
    (H_{\text{spin}})_{nn} &= J \, \hat{\vb{e}}_n \cdot \bm{\sigma}, 
    \\[0.2cm]
    \hat{\vb{e}}_{\vb{r}_1} &= \tfrac{1}
    {\sqrt{2}} (-1,1,0)^T,
    \\
    \hat{\vb{e}}_{\vb{r}_2} &= \tfrac{1}
    {\sqrt{2}} (1,1,0)^T,
    \\
    \hat{\vb{e}}_{\vb{r}'_1} &= \tfrac{1}
    {\sqrt{2}} (1,-1,0)^T,
    \\
    \hat{\vb{e}}_{\vb{r}'_2} &= \tfrac{1}{\sqrt{2}} (-1,-1,0)^T,
\end{align}
where $J \in \mathbb{R}$ defines the coupling strength. 
The explicit symmetry representations in the basis of Eq.~\eqref{Eq_fullTBham} are
\begin{align}
    M_y &= \begin{pmatrix}
    0 & 1 & 0 & 0 \\ 
    1 & 0 & 0 & 0 \\ 
    0 & 0 & 0 & 1 \\ 
    0 & 0 & 1 & 0 \\ 
    \end{pmatrix}
    \otimes
    i \sigma_y
    \label{Eq_EndMatter_MyRep}
    \\
    C_{3z} &= \begin{pmatrix}
    1 & 0 & 0 & 0 \\ 
    0 & 1 & 0 & 0 \\ 
    0 & 0 & 1 & 0 \\ 
    0 & 0 & 0 & 1 \\ 
    \end{pmatrix}
    \otimes
    \sigma_0
    \\
    \widetilde{\mathcal{T}} &= \begin{pmatrix}
    0 & 0 & 1 & 0 \\ 
    0 & 0 & 0 & 1 \\ 
    1 & 0 & 0 & 0 \\ 
    0 & 1 & 0 & 0 \\ 
    \end{pmatrix}
    \otimes
    i \sigma_y K, \text{ and}
    \\
    \widetilde{C}^s &= \begin{pmatrix}
    0 & 0 & 1 & 0 \\ 
    0 & 0 & 0 & 1 \\ 
    1 & 0 & 0 & 0 \\ 
    0 & 1 & 0 & 0 \\ 
    \end{pmatrix}
    \otimes
    i \sigma_z
    \label{Eq_EndMatter_spinscrew}.
\end{align}

A 3D variant of the model can be obtained by replacing $t_2 \rightarrow 2 t_2 \cos(\frac{\vb{a}_3}{2} \cdot \vb{k})$.
Hereby, the layer exchange operation in $\widetilde{C}^s$ and  $\widetilde{\mathcal{T}}$ is replaced by a translation $\vb{t}_{1/2}$.
The symmetry representations for the 3D model are still given by Eqs.~\eqref{Eq_EndMatter_MyRep} --\eqref{Eq_EndMatter_spinscrew}, except that $\widetilde{\mathcal{T}}$ and $\widetilde{C}^s$ are multiplied by $\mathrm{e}^{i k_z/2}$ to account for the translation representation of $\vb{t}_{1/2}$.
At $k_z = \pi$, the Kramers theorem does not apply to $\widetilde{\mathcal{T}}$. 
Nevertheless, the anticommutation of $M_y$ and $\widetilde{C}^s$ enforces three intersecting nodal planes with $f$-wave splitting symmetry.

\textit{Low-energy model} \label{EndMatter_LowEnergyModel} -- The symmetry-allowed terms in the expansion of Eq.~\eqref{Eq_lowEHam} up to $O(k_i^5)$ are
\begin{align}
    &E_0^{(1)} k^2 \tilde{\sigma}_0, \,\, E_0^{(2)} k^4 \tilde{\sigma}_0,  \\
    &\Delta_s^{(1)} k_y (3 k_x^2 - k_y^2) \tilde{\sigma}_z, \,\, \Delta_s^{(2)} k^2 k_y (3 k_x^2 - k_y^2) \tilde{\sigma}_z,
\end{align}
with $k^2 = k_x^2 + k_y^2$. 
The prefactors $E_0^{(1)}, E_0^{(2)}, \Delta_s^{(1)}, \Delta_s^{(2)} \in \mathbb{R}$ were fitted to the tight-binding band structure, see Fig.~\ref{Fig2}(b).
The first symmetry-allowed deviation from $f$-wave symmetry arises to order $O(k^9)$ with $l=9$ nodal lines, which is of negligible size.

To obtain the analytic expansion of the $8 \times 8$ model $H(\vb{k}) $, we block-diagonalize the Hamiltonian using the eigenstates of $\widetilde{C}^s$. 
For the resulting $4\times4$ blocks, we calculate the eigenstates setting $t_1 = t_2$ and $\vb{k} = \vb{0}$.
We project $H(\vb{k})$ onto these eigenstates to obtain the final $2 \times 2$-block structure, assuming that the degenerate band pairs are well-separated.
By expanding the $2 \times 2$ blocks in $\vb{k}$, we recover the symmetry-derived low-energy model Eq.~\eqref{Eq_lowEHam} with coefficients Eqs.~\eqref{Eq_AnalyticCoefficients}.

\textit{Boltzmann transport} \label{EndMatter_Boltzmanntransport} -- 
The introduced Hamiltonians exhibit no Berry curvature, and thus we focus our transport considerations on intraband effects captured qualitatively by Boltzmann transport \cite{Zelse2017}. 
In the relaxation time approximation, the conductivity, spin conductivity, and Edelstein susceptibility are given by \cite{johansson2024theory, Cano2024, Ezawa2025higherOrderResponses}
\begin{align}
    \sigma_{xx} &= 
    - e^2 \tau  \sum_n   \int  \frac{\text{d} k^2}{(2\pi)^2} \, \frac{\partial f^{(0)}(E_n)}{\partial E}  v_{n,x}(\vb{k})^2,
    \label{Eq_BoltzSigXX}
    \\
    \sigma^s_{xx} &= 
    e \tau  \sum_n   \int  \frac{\text{d} k^2}{(2\pi)^2} \, \frac{\partial f^{(0)}(E_n)}{\partial E}  j^s_{n,x}(\vb{k}) v_{n,x}(\vb{k}),
    \label{Eq_BoltzSpinSigXX}
    \\
    \chi_{zy}  &=  e \tau  \sum_n   \int  \frac{\text{d} k_y}{2\pi} \, \frac{\partial f^{(0)}(E^x_n)}{\partial E}  s_{z,n}(k_y) v_{n,y}(k_y) ,
    \label{Eq_EdelsteinFormula}
\end{align}
respectively, where $e$ is the absolute value of the elementary charge,  $\tau$ is the scattering time, the lattice constant is omitted, $\hbar = 1$, $f^{(0)}(E)$ is the equilibrium Fermi-Dirac distribution, $v_{n,i}(\vb{k})= \frac{\partial E_n}{\partial k_i}$ is the group velocity, $s_{z,n}$ is spin polarization of band $n$, and spin current expectation value is defined as $j^s_{n,x}(\vb{k}) = \langle \psi_n(\vb{k}) \vert \tfrac{1}{2} \{s_x, \tfrac{\partial H}{\partial k_x} \} \vert \psi_n(\vb{k}) \rangle$.
Note, for the Edelstein susceptibility $\chi_{zy}$, the expression is given for the edge theory or ribbon tight-binding model, i.e., $E^x_n$ is the $n$th energy band of the $x$-terminated ribbon, and the velocities and spin in Eq.~\eqref{Eq_EdelsteinFormula} are defined accordingly.

For a low-energy model, $s_{z,n}(k_y)$ is piecewise constant in $k_y$.
At $T=0$, the integration simplifies to a sum over the momenta $k$ of the Fermi surface for band $n$ 
\begin{align}
    \chi_{zy}  =  - \frac{e \tau}{2\pi \hbar}  \sum_{n} \sum_{k \in \text{FS}}  \text{sgn}(v_{n,y}(k)) s_{z,n}(k).
    \label{Eq_EdelsteinAnalytic}
\end{align}
For a relaxation time with $\tfrac{e \tau}{\hbar} = 100/$V and a lattice constant of $5$~\AA\ the numerical values in Figs.~\ref{Fig3}(d,e) are comparable to the bulk Edelstein effect in $p$-wave magnets \cite{chakraborty2025highly}.

%%%%%%%%%%%%%%%%%%%%%%%%%%%%%%%%%%%%%%%%%%%%%%%%%%%%
%%%%%%%%%%%%%%%%%%%%%%%%%%%%%%%%%%%%%%%%%%%%%%%%%%%%

\textit{Dirac points and edge states} -- The symmetries, $C_{3z}$ and spinful $M_y$,  protect Dirac points at K and K' with $\pi$-Berry phase, which are equivalent to the Dirac points in graphene \cite{castro2009electronic}.
Here, the group structure related to $C_{3z}$ and $M_y$ is identical with and without spin, thus the Dirac points persist in the magnetic structure.
While such Dirac points also arise in the $f$-wave model of Ref.~\cite{uchino2025analysis}, they are not generally enforced in $f$-wave magnets  \cite{zhu2025floquet, huang2025light}.

For the $x$ termination, i.e, the armchair termination, there are no edge states as K and K' are mapped onto the same momentum.
In contrast to graphene, $\widetilde{C}^s$ protects with open boundary conditions two linear crossings (Fig.~\ref{Fig3}(c)).
These crossings exhibit $p$-wave splitting at $E=\pm 3.4$ corresponding to a local maximum in Fig.~\ref{Fig3}(d). 
The Dirac points are well separated in energy from the dominant contribution to the $f$-wave Edelstein effect and are therefore not essential for it.

\clearpage

%\begin{widetext}
%\section*{Supplementary Information}
%
%\end{widetext}

\end{document}